\documentclass[acus]{pac99}

\usepackage{epsfig}

\newcommand{\bit}{\begin{Itemize}}
\newcommand{\eit}{\end{Itemize}}

\begin{document}
\title{MUON COLLIDERS: THE ULTIMATE NEUTRINO BEAMLINES}
\author{ Bruce J. King, Brookhaven National Laboratory $^1$ }
\maketitle

\begin{abstract}

  It is shown that muon decays in straight sections of muon collider rings
will naturally produce highly collimated neutrino beams that can be several
orders of magnitude stronger than the beams at existing accelerators.
We discuss possible experimental setups and give a very brief overview of
the physics potential from such beamlines.
Formulae are given for the neutrino event rates
at both short and long baseline neutrino experiments in
these beams.

\end{abstract}

\footnotetext[1]{
web page: http://pubweb.bnl.gov/people/bking/,
email: bking@bnl.gov.
This work was performed under the auspices of
the U.S. Department of Energy under contract no. DE-AC02-98CH10886. }

\section{Introduction}
\label{sec-intro}

 Recent feasibility studies and design work for muon
colliders~[1] has begun to focus attention on the
exciting physics possibilities from the uniquely intense neutrino beams
at proposed muon storage rings (muSR) that will use muon collider
technology to produce and store large muon currents.
Such muSR include both storage rings dedicated to neutrino physics
and the parasitic use of the accelerator and collider rings of muon
colliders themselves.

\section{Potential for Neutrino Physics}

 The neutrino beamlines can be used for two sorts of
neutrino experiments:
\begin{enumerate}
  \item  short baseline (SB) experiments, where
the detector is placed close to the neutrino source to obtain
the most intense beam possible and hence gather very high event
statistics of neutrino interactions.
  \item  long baseline (LB) experiments, where a very
massive neutrino detector is placed far away
from the neutrino source, deliberately sacrificing event
rate in order to study baseline-dependent properties of the
neutrinos and, in particular, whether there are ``flavor
oscillations'' in the types of neutrinos composing the
beam.
\end{enumerate} 

 The large muon currents and tight collimation of the neutrinos results
in extremely intense beams -- intense enough even to constitute a potential
off-site radiation hazard~[2]. This gives several advantages
over the neutrino beams produced today from pion decay at accelerator
beamlines:
\begin{enumerate}
 \item  event statistics for short-baseline (SB) experiments that
might be three or more orders-of-magnitude larger than in today's
high-rate neutrino experiments
 \item  both higher
statistics and longer baselines for long baseline (LB) experiments
 \item  extremely well understood and pure two-component beams with accurately
predictable energy spectra, angular divergences and intensities
 \item  the first high flux electron-neutrino and electron-antineutrino
beams at high energies
 \item  the possibility of tuning the flavor composition of the
beam by varying the polarization direction of the muons.
\end{enumerate}

 Recent evidence of neutrino oscillations from several different
experiments has resulted in much recent interest in LB neutrino
oscillation experiments. LB experiments at muSR
have the potential~[3,4,5] to either convincingly refute
or precisely characterize much of today's evidence for neutrino
oscillations.

  SB experiment have a different physics motivation, involving
several precision measurements that can contribute mainly to our
understanding of the interactions between elementary particles.
In this respect, they have the potential~[3,4,5] to make
important measurements have comparable value to (and are complementary
to) some of the best precision measurements at today's colliders:
LEP, SLC, CLEO, Tevatron, HERA and the B factories that are now
coming on-line.

\begin{figure}[hbt]
\centering
\epsfig{file=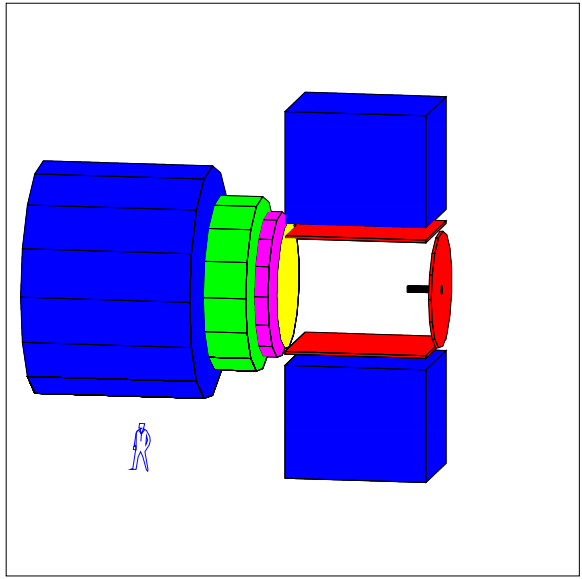, width=3.0in}
\caption{Example of a general purpose SB neutrino detector. A human figure in
the lower left corner illustrates its size. The neutrino target is the small
horizontal cylinder at mid-height on the right hand side of the detector. Its
radial extent corresponds roughly to the radial spread of the neutrino pencil
beam, which is incident from the right hand side. Further details are given
in reference~[3].}
\label{nudet}
\end{figure}

  The huge increase in beam intensity allows the use of novel
high-performance SB neutrino detectors that are capable of
much better reconstructing neutrino events than today's huge
high-rate neutrino detector targets. An example of such a
detector~[3] is shown in figure~\ref{nudet}.

\section{Neutrino Production and Event Rates}

  Neutrinos are emitted from the decay of muons in the collider ring:
\begin{eqnarray}
\mu^- & \rightarrow & \nu_\mu + \overline{\nu_{\rm e}} + {\rm e}^-,
                                             \nonumber \\
\mu^+ & \rightarrow & \overline{\nu_\mu} + \nu_{\rm e} + {\rm e}^+.
                                                 \label{eq:nuprod}
\end{eqnarray}

   The thin pencil beams of neutrinos for experiments will be produced
from long straight sections in either the collider ring or a
ring dedicated to neutrino physics. From relativistic kinematics, the
forward hemisphere in the muon rest frame will be boosted, in the lab
frame, into a narrow cone with a characteristic opening half-angle,
$\theta_\nu$,
given in obvious notation by
\begin{equation}
\theta_\nu \simeq \sin \theta_\nu = 1/\gamma =
\frac{m_\mu}{E_\mu} \simeq \frac{10^{-4}}{E_\mu ({\rm TeV})}.
                                                   \label{eq:thetanu}
\end{equation}

 The decays of muons into neutrinos are very well understood
and it is possible~[6] to derive simple, approximate quantitative
expressions for the neutrino fluences at experiments and the corresponding
experimental event rates in a given target.

 The neutrino targets at SB experiments might reasonably cover the
boosted forward hemisphere of the neutrino beam. Besides the beam
properties, the event rate depends on the depth of the neutrino
target, $l$, in units of gm.cm$^{-2}$ and on the number of years
of data-taking, Y. Thus, a convenient benchmark for the size of neutrino
event samples the beam can produce is ${\rm  N^{sb}}$, defined as
\begin{eqnarray}
{\rm no.\;of\;\nu\; events} =
                 {\rm  N^{sb}[events/yr/(gm.cm^{-2}])
                 \times }
                 \nonumber \\
l{\rm [gm.cm^{-2}] \times Y },
              \label{eq:beam_nsbdef}
\end{eqnarray}

  Ref~[6] derives the following numerical expression for
${\rm  N^{sb}}$ in SB detectors satisfying a couple of reasonable
assumptions:
\begin{eqnarray}
{\rm  N^{sb}[events/yr/(gm.cm^{-2}])} \simeq
       2 \times 10^{-15} \times
           \nonumber \\
 E_\mu[GeV] \times
       N^{ss}_\mu[yr^{-1}]
              \label{eq:beam_nsb},
\end{eqnarray}
where the parameter $N^{ss}_\mu$ is the number of forward-going muons
decaying in the production straight section per year.

Similarly, the number of LB events depends on the neutrino beam,
the baseline distance, L, from neutrino source to detector and
the detector mass (L) through:
\begin{eqnarray}
{\rm no.\;of\;\nu\; events} =
             \nonumber \\
                 {\rm  N^{lb}[events/yr/(kg.km^{-2})]
                 \times
 \frac{M[kg]}{(L[km])^2} \times Y  }
              \label{eq:beam_nlbdef},
\end{eqnarray}
with
\begin{eqnarray}
{\rm  N^{lb}[events/yr/(kg.km^{-2})]} \simeq
                \nonumber \\
\frac{ 1.6 \times 10^{-20}
      \times N^{ss}_\mu[yr^{-1}] \times \times (E_\mu[GeV])^3 }
     {(\gamma \cdot \delta \theta)^2}
              \label{eq:beam_nlb}.
\end{eqnarray}
(The parameter, $\gamma \cdot \delta \theta$ has
value 1 except for neutrino beams from muon beams with large angular
divergences.)

 Reference~[6] also tabulates the expected event rates for
plausible muon collider scenarios. These predictions are reproduced
as table~\ref{tab:beam_specs}. Extraordinary event samples of
$10^{10}$ interactions appear plausible for SB experiments at muSR
with muon energies of order 100 GeV or above.

\section{Conclusions}
\label{sec-conc}

 In conclusion, the essentially {\em free} neutrino beamlines
at muon colliders and the beams from other muSR using muon
collider technology could provide happy prospects for the
future of neutrino physics.

\section{references}

\noindent[1] The Muon Collider Collaboration,
``Status of Muon Collider Research
and Development and Future Plans'', to be submitted to Phys. Rev. E.

\noindent [2] B.J. King, ``Potential Hazards from Neutrino Radiation
at Muon Colliders'', these proceedings.

\noindent [3] B.J. King,
  ``Neutrino Physics at a Muon Collider'',
 Proc. Workshop on Physics at the First Muon Collider
 and Front End of a Muon Collider, Fermilab, November 6-9, 1997.
    Available at http://pubweb.bnl.gov/people/bking/. 

\noindent [4] B.J. King,
  ``Neutrino Physics at Muon Colliders'',
 Proc. 4th Int. Conf. on Workshop on the Physics Potential and
Development of Muon Colliders, San Francisco, December 1997,
ed. David B. Cline.
    Available at http://pubweb.bnl.gov/people/bking/. 

\noindent [5] B.J. King, ``Muon Colliders: New Prospects for Precision
Physics and the High Energy Frontier'',
BNL CAP-224-MUON-98C,
Proc. Latin Am. Symp.
on HEP, April 8-11,1998, San Juan, Puerto Rico, Ed. J.F. Nieves.
Available at http://pubweb.bnl.gov/people/bking/. 

\noindent [6] I. Bigi {\it et al.}
    ``An Overview of the Potential for Neutrino Physics at Future
      Muon Collider Complexes''. Work in progress. Contact B.J. King,
      email: bking@bnl.gov .

\newpage

\begin{table}[htb!]
\begin{center}
\caption{Predicted neutrino fluxes and event rates for muSR's
or muon colliders at several energies.}
\begin{tabular}{|r|cccccc|}
\hline
\multicolumn{1}{|c|}{ {\bf muon energy, ${\rm E_\mu}$} } &
                      2 GeV & 20 GeV  & 50 GeV & 175 GeV & 500 GeV & 5 TeV  \\
\hline
($\mu^-$ or) $\mu^+$/year,${\rm N_\mu[10^{20}}]$ &
                                       3.0 & 3.0 & 6.0 & 6.0 & 3.2 & 3.6 \\
fract. str. sect. length, ${\rm f_{ss}}$ & 
                           0.40 & 0.30 & 0.12 & 0.12 & 0.12 & 0.03 \\
$\nu$ ang. diverg., $\gamma \cdot \delta \theta_\nu$ &
                                          1 & 1 & 1 & 1 & 10 & 1 \\
time to beam dump,
          ${\rm t_D} [\gamma \tau_\mu]$ & no dump & no dump & no dump
                                                & no dump & 0.5 & no dump \\
$N^{sb}$[evt/yr/(${\rm g.cm^{-2}})$]
                                             & $5.1 \times 10^5$ 
                                             & $3.8 \times 10^6$
                                             & $6.5 \times 10^6$
                                             & $2.7 \times 10^7$
                                             & $2.3 \times 10^7$
                                             & $1.0 \times 10^8$ \\
$l \times Y$[${\rm gm/cm^2}$] for $10^{10}$ events
                                             & $2.0 \times 10^4$
                                             & 2600
                                             & 1500
                                             & 370
                                             & 430
                                             & 100 \\
$N^{lb}$[evt/yr/(${\rm kg.km^{-2}}$]
                                              & 16
                                              & $1.2 \times 10^4$
                                              & $1.4 \times 10^5$
                                              & $6.2 \times 10^6$
                                              & $5.0 \times 10^5$
                                              & $2.2 \times 10^{10}$ \\
\hline
\end{tabular}
\label{tab:beam_specs}
\end{center}
\end{table}

\end{document}